\documentclass[aip,rsi,reprint,graphicx]{revtex4-1} 
\usepackage{graphicx} 
\usepackage{epstopdf}
\usepackage{amsmath}
\usepackage{mathtools}
\usepackage{amssymb}

\draft 

\begin{document}


\title{Open Microwave Cavity for use in a Purcell Enhancement Cooling Scheme } 



\author{N. Evetts}
\email[]{nevetts@phas.ubc.ca}

\affiliation{Department of Physics and Astronomy, University of British Columbia, Vancouver BC, Canada V6T 1Z1}

\author{I. Martens}
\author{D. Bizzotto}
\affiliation{Department of Chemistry, AMPEL, University of British Columbia, Vancouver BC, Canada V6T 1Z1}

\author{D. Longuevergne}
\affiliation{Institut de Physique Nucl\'eaire, CNRS/IN2P3, Universit\'e Paris-Sud, UMR 8608, 91406 Orsay, France}

\author{W. N. Hardy}
\affiliation{Department of Physics and Astronomy, University of British Columbia, Vancouver BC, Canada V6T 1Z1}


\date{\today}

\begin{abstract}
A microwave cavity is described which can be used to cool lepton plasmas for potential use in synthesis of antihydrogen. The cooling scheme is an incarnation of the Purcell Effect: when plasmas are coupled to a microwave cavity, the plasma cooling rate is resonantly enhanced through increased spontaneous emission of cyclotron radiation. The cavity forms a three electrode section of a Penning-Malmberg trap and has a bulged cylindrical geometry with open ends aligned with the magnetic trapping axis. This allows plasmas to be injected and removed from the cavity without the need for moving parts while maintaining high quality factors for resonant modes. The cavity includes unique surface preparations for adjusting the cavity quality factor and achieving anti-static shielding using thin layers of nichrome and colloidal graphite respectively. Geometric design considerations for a cavity with strong cooling power and low equilibrium plasma temperatures are discussed. Cavities of this weak-bulge design will be applicable to many situations where an open geometry is required.
\end{abstract}


\maketitle 

\section{Introduction}

\begin{figure}
\includegraphics[width=\columnwidth]{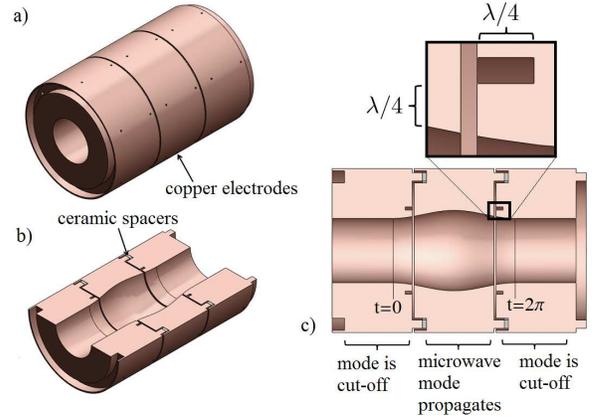}
\caption{\label{fig:cavity} A microwave ``bulge" cavity is formed from three copper electrodes electrostatically isolated by ceramic spacers. The inner surface of the cavity is parametrized according to equations \ref{eq:Param1}-\ref{eq:Param3}. The inset of c) shows a typical microwave ``choke": a quarter wave structure which prevents radiation leakage through the gap between electrodes.}%
\end{figure}

Resonant electromagnetic cavities have widespread applications in science and engineering. 
We present a microwave cavity within the context of its application: a new cooling technique for lepton plasmas representing an incarnation of the Purcell Effect \citep{Purcell}. The technique holds promise for improving antihydrogen synthesis schemes. While this is the only application discussed, we emphasize that this type of cavity could be useful in any situation where an open geometry is required. The resonator geometry (figure \ref{fig:cavity}) resembles that used in gyrotrons \citep{Nusinovich} and at radio frequency particle accelerators \cite{Padamsee}. The cavity features to be emphasized here are the large (relative to the radiation wavelength) axial ``windows" and the slightness of the central ``bulge" region where the inner diameter changes gradually. 

\begin{figure}
\includegraphics[width=\columnwidth]{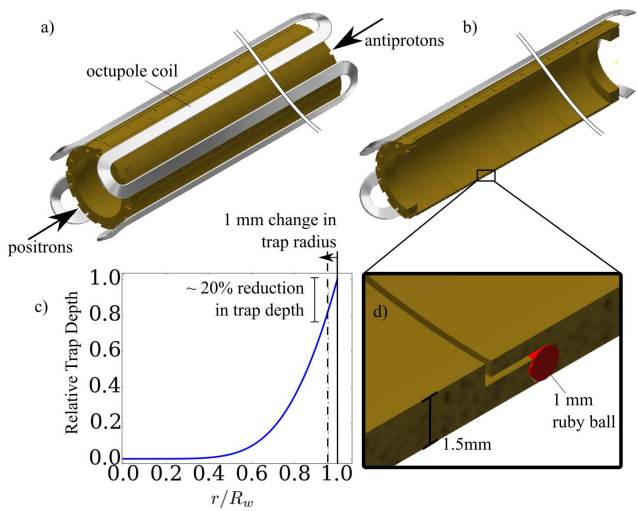}
\caption{\label{fig:alpha} a), b), and d) shows the ALPHA Penning trap and magnetic trap in various views. c) shows the magnetic field as a function of radius (normalized to the inner radius of the electrodes $R_w = 22.275$ mm). A small reduction in $R_w$ greatly reduces the depth of the trap.}%
\end{figure}

For our application, the slightness of this bulge is particularly important. The antihydrogen magnetic trap at ALPHA \citep{Amole:apparatus_etal}, for example, consists of an octupole coil tightly wound near the outer face of a Penning trap (figure \ref{fig:alpha}). The choice of an octupole coil over some other multipole represents a compromise between the competing needs to produce strong magnetic fields which radially confine antihydrogen, while maintaining a homogeneous field at the trap centre where non-neutral plasmas are confined and manipulated (this issue is summarized elsewhere \citep{Bertsche_etal}). In the ALPHA experiment, typical diameters of plasmas are 1 mm, for which broadening of the cyclotron frequency by the octupole coil is negligible.


The Penning trap has a central bore of 44.55 mm with electrodes having a wall thickness of only 1.5 mm. The electrode wall thickness has been minimized in order to maximize the depth of the magnetic trap. Due to the sharply rising potential well created by the magnetic field from the octupole coil, an increase in the thickness of the Penning trap electrode walls of only 1 mm would result in a $\sim 20 \%$ reduction in the antihydrogen trapping rate. The task of designing a cavity resonator without changing the Penning trap geometry by more than a millimetre presents a significant challenge.

We show that a bulge cavity such as we are presenting can create localized, high-$Q$ resonances with an inner diameter change of only 2.5 mm. Provided that the extra cooling of the positrons can increase the trappable antihydrogen population such that losses due to a lower trap depth are offset, this technique could increase the number of antiatoms available for experimentation at ALPHA. While experiments utilizing electron plasmas are currently under way to probe this question, preliminary data\citep{Povilus:PRL,Povilus} suggest that the plasma cooling rate can be enhanced by a factor of $\sim 10$. These exploratory experiments do not involve magnetic traps and accordingly do not require thin electrodes. It is the \textit{slightness} of the inner bulge that is relevant.

\subsection{Purcell Enhancement}
\label{sec:Purcell}

When plasmas are coupled to a microwave cavity resonant at the cyclotron frequency, the plasma cooling rate is enhanced through increased spontaneous emission of cyclotron radiation. By multiplying the free space cyclotron emission rate,

\begin{equation}
\Gamma_{free \, space} = \frac{2}{3} \times \frac{e^2 \omega_c^2}{3 \pi \epsilon_o c^3 m_e},
\label{eq:RateFree}
\end{equation} 
by Purcell's enhancement factor $\left( \frac{3 g Q \lambda^3}{4 \pi^2 V} \right)$, one obtains an expression for the enhanced emission rate for a single electron in cyclotron motion

\begin{align}
\Gamma_{enhanced} &= \frac{4 g Q e^2}{3 \epsilon_o m_e \omega_c V_{eff}}  \\
&= \frac{4 g Q e^2}{3 \epsilon_o m_e \omega_c} \frac{E_T^2(\vec{x})}{\int_V \! E^2(\vec{x}') \, \mathrm{d}x'^3 }. \label{eq:RateChi}
\end{align}
Above, $Q$, $V_{eff}$, $g$, and $E_T$ are the resonance quality factor, effective volume, degeneracy, and transverse electric field; $e$ is the elementary charge; $m_e$ is the electron mass; $\vec{x}$ is the position of the plasma; and $\omega_c$ is the cyclotron angular frequency. The pre-factor of 2/3 in equation \ref{eq:RateFree} accounts for equilibration of the two cyclotron degrees of freedom with the third, axial, degree of freedom. In equation \ref{eq:RateChi} we have replaced $1/V_{eff}$ with a ``fill factor" $\chi = E_T^2(\vec{x})/\int_V \! E^2(\vec{x}') \, \mathrm{d}x'^3 $ necessary for any cavity with non-homogeneous electric fields ($\vec{E}$). This factor quantifies the strength of the coupling between the emitter and the cavity, and can be motivated either classically \citep{Evetts:Thesis} or from cavity quantum electrodynamics \citep{Boroditsky}. The rate enhancement resulting from a cavity therefore depends on the optimization of a collection of parameters: resonance degeneracy, cyclotron frequency, quality factor, and fill factor ($\chi$). Conversely, the cooling rate should be suppressed when the plasma is located at a transverse electric field null $E_{T}(\vec{x})=0$, or if the density of states for photons is small (for example, when the cyclotron frequency is very different from a cavity resonant frequency).

Previous work has described cavity enhanced (and suppressed) spontaneous emission by including the electric fields of microwave cavity resonances in the equations of motion for a charged particle in a Penning trap.\citep{BrownPRA:1985} This approach is considerably more complex and generally more applicable to simple cavity geometries. It used the method of images to obtain solutions for both the electric fields inside the cavity and the surface currents in the electrode walls. An alternative theoretical description is provided by O'Niel \citep{ONiel} which predicts that a plasma cyclotron frequency distribution $\Delta \omega_c \sim \omega_o/Q$ would optimize the cooling effect. The measured width of the cyclotron resonance at ALPHA is typically a few tens of MHz \citep{Friesen_etal} to be compared with cavity resonance widths of 10 - 30 MHz (from table \ref{tab:Sim}).

\subsection{Cooling of $N$ particles}

The enhanced cooling rate given in equation \ref{eq:RateChi} is derived for the case of a single particle coupled to a resonant cavity. In order to be useful for antihydrogen synthesis, however, millions of positrons must be cooled. In the limit of large lepton number ($N\gg1$), the refrigerative power of the cavity will become ineffective when distributed over too many particles. To first order we expect the maximum number of particles which our cavity is able to cool to go as 

\begin{equation}
N_{max} \sim \frac{\Gamma_{cavity}}{\Gamma_{enhanced}}
\label{eq:Overload}
\end{equation}
where $\Gamma_{cavity} = \omega_c/Q$. Here we have used the single particle cooling rate as an approximation for the average $N_{max}$ particle cooling rate. In this limit, we have 

\begin{equation}
N_{max} \propto \frac{1}{Q^2}.
\end{equation} 

While studying shifts of the cyclotron frequency resulting from cavity effects, other authors have suggested that the relevant ``overload" number (equation \ref{eq:Overload}) is given by \citep{Tan:1993}

\begin{equation}
N^2_{max} \sim \frac{\Gamma_{cavity}}{\Gamma_{enhanced}},
\label{eq:OverloadSync}
\end{equation}
which implies phase synchronization of the cyclotron motion. In either case, lowering the cavity $Q$ will raise $N_{max}$.

Therefore, while we desire a high $Q$ for a high cooling rate, we desire a low $Q$ in order to be able to cool large numbers of leptons, and a compromise value is needed. As an example, if one uses the expected $Q \sim 10^4$ for a room temperature copper cavity in equation \ref{eq:RateChi} (with the TE$_{131}$ $f$ and $\chi$ from table \ref{tab:Sim}), one obtains a cooling time of order milliseconds, which represents an improvement over the free space value (equation \ref{eq:RateFree}) by a factor $\sim$ 1000. However, such a high $Q$ may severely limit the number of particles that can be cooled. Section \ref{sec:surface} describes efforts to lower the $Q$ by a factor of 10 or so, in order to increase $N_{max}$ while still maintaining substantial improvement in the global cooling rate.

\section{Cavity Design}

The cavity resembles a typical cylindrical cavity except that the usual boundary conditions at the axial end faces of such a resonator have been removed. Instead we create a central ``bulge" region where cut-off frequencies of typical waveguide modes are lower than in the surrounding regions. To either side of this bulge, therefore, microwaves are below cut-off and thus do not propagate.

The segmented nature of the resonator is dictated by the need to hold regions of the cavity at different DC voltages as is standard in any Penning trap apparatus. The small gaps between electrodes allow microwave radiation to leak radially, which affects the electric field structure and $Q$ of the trapped resonances. To counteract this effect we incorporate chokes \citep{Ragan} into this small gap (see the inset of figure \ref{fig:cavity} (c)). 

\begin{figure}
\includegraphics[width=\columnwidth]{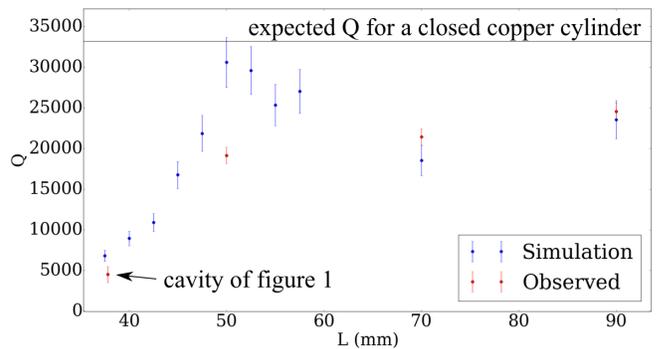}
\caption{\label{fig:QStudy} The dependence of the cavity $Q$ on the length over which the bulge occurs. The red points at $L=50$ mm, 70 mm, and 90 mm represent cavities which have been machined from a single piece of aluminium. The results can be compared to the expected $Q$ for the TE$_{131}$ resonance of a right circular cylinder of dimensions $L = 40$ mm and $D = 25$ mm (horizontal line).}%
\end{figure}

For the cavity of figure \ref{fig:cavity}, small amounts of radiation leak axially down the open ends. This is inferred by both simulation and direct measurement of the $Q$. The results were $Q=6800$ and $Q=4500$ respectively for the TE$_{131}$ mode of a copper ``bulge" cavity at room temperature (to be compared to an expected $Q \sim 10^4$). Leakage is an undesirable mechanism for lowering the $Q$. The presence of leakage allows radiation to couple into the cavity from other, possibly warm, parts of the experiment causing unwanted heating of plasmas. This subtlety has been noted in previous Purcell Effect experiments \citep{Goy}.

The smooth surface which forms the belly of the bulge cavity is parametrized by

\begin{align}
r(t) &= -\frac{\Delta R}{2} \cos(t) + R_{mean} \label{eq:Param1} \\
z(t) &= \frac{L}{2 \pi} t + b \label{eq:Param2} \\
t & \in [0,2 \pi]  \label{eq:Param3}
\end{align}
where, for our case $\Delta R =2.5$ mm, $L = 37.8$ mm, $R_{mean} = 11.25$ mm, and $b$ is a constant which places the maximum $r(t)$ at the center of the three electrode cavity. It is clear that lower $Q$ values will result from more abrupt bulges which cause conversion to modes that propagate in both end regions. This loss mechanism can be greatly reduced in future designs by using a bulge that transitions more gradually from the narrower end regions to the central (but still slightly) bulged region. 

Figure \ref{fig:QStudy} shows the $Q$ dependence on the ``smoothness" of the  bulge. There we compare simulation results to both the expected $Q$ of a closed resonator, as well as $Q$'s measured in longer bulge cavities machined from aluminium. The aluminium cavities were not segmented into multiple electrodes and their $Q$'s are scaled up by factors of $\sim 15 \%$. This is a result of modifying the ohmic contribution to the $Q$ by a factor  $\sqrt{\sigma_{\mathrm{Cu}} / \sigma_{\mathrm{Al}}} \sim 1.3$. Simulation results using both finite and zero resistivity metallic walls allow estimation of the ohmic contribution according to $1/Q = 1/Q_{\Omega} + 1/Q_{leakage}$. The simulated $Q_{leakage}$ are $5 \times 10^4$, $4 \times 10^4$, and $8 \times 10^4$ for single-electrode cavities with $L = $50, 70, and 90 mm respectively. For the cavity of figure \ref{fig:cavity}, $Q_{leakage}$ is about 9000.  A relatively small increase in the length of the bulge (from 37.8 mm to 50 mm) gives more than a four-fold increase in the measured $Q$. 

\begin{figure}
\includegraphics[width=\columnwidth]{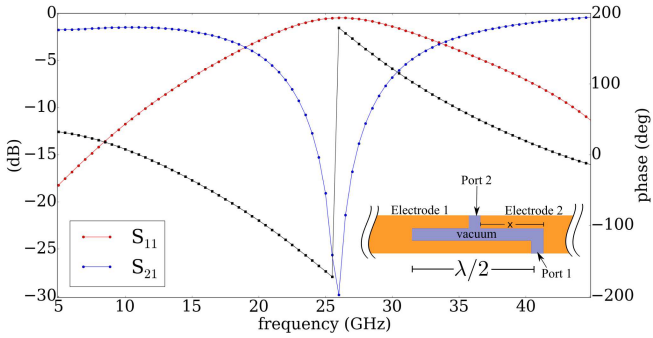}
\caption{\label{fig:L2Choke} A ``choke" of length $\lambda/2 \sim 5.5$ mm, and width $< \lambda/4$. A simulation result shows good reflection near the design frequency of $\sim 27$ GHz. The reflection phase is shown in black. Its value at resonance can be modified by fine tuning the horizontal distance ($x$) between port 1 and 2. Here we have used $x=2.25$ mm.}%
\end{figure}

In order to obtain these higher $Q$'s, two of the four coupling holes which were drilled at the centre of the aluminium cavities (similar to figure \ref{fig:VNA}) had to be filled. Our conclusion is that these holes act as scattering centres and serve to increase radiation leakage from the cavity.

Preliminary simulations of bulge cavities comprised of many, thin-walled electrodes with the current inter-electrode geometry in use at ALPHA (as in figure \ref{fig:alpha}d) have been performed. Results suggest that cavity resonances can easily be created, but with $Q$'s ($\sim 10^3$) which are affected by radiation loss. Methods to alleviative radiation loss from potential cavities in the ALPHA Penning trap might include designs which place cavity electric field nulls at the axial locations of the gaps, or those which incorporate thin, $\lambda/2$ style chokes into the inter-electrode geometry. We have simulated such a choke (see figure \ref{fig:L2Choke}), although it has not yet been built or tested experimentally.

\subsection{Simulations}

\begin{figure}
\includegraphics[width=\columnwidth]{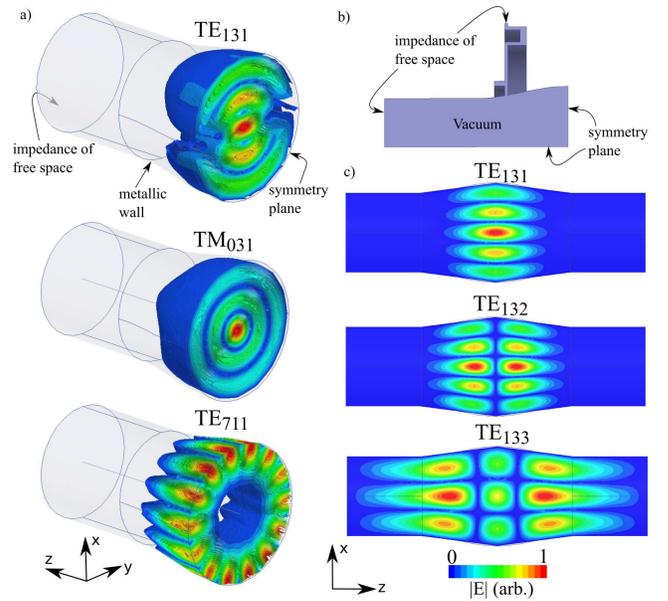}
\caption{\label{fig:131} 
a) and c) Show cross-sectional views of the cavity overlaid with the simulated electric field structure for various resonant modes. b) shows the cavity geometry and simulated boundary conditions producing results given in table \ref{tab:Sim}. Boundaries not labelled in b) are metallic. Simulated fields in a) and c) use different symmetry conditions and omit the choke and inter-electrode structure in order to more clearly display the cavity fields (which change negligibly for this purpose). The dimensions for this cavity are given in the text below equation \ref{eq:Param3}. }%
\end{figure}

The complicated geometry of this resonator makes analytical calculations of the field patterns, resonant frequencies, and quality factors impractical. We therefore resort to numerical results and rely on the commercial simulation software, HFSS \citep{HFSS}, which uses finite element methods. Various cavity resonances are illustrated in figure \ref{fig:131}. Other resonances shown to be able to cool plasmas via the Purcell Effect\citep{Povilus:PRL} are detailed in table \ref{tab:Sim}. Note that, although many TE and TM modes are resonant in this cavity, in order to achieve strong cooling (via a high $\chi$) for plasmas on the cylindrical axis of the trap, we only consider TE$_{1lm}$ modes which have strong on-axis electric fields that are polarized perpendicular to the cylindrical axis.

\begin{table}
\caption[Microwave resonances of the bulge cavity.]{ \label{tab:Sim} A summary of the simulated cavity resonances. The simulation uses a conductivity similar to nichrome ($\sigma = 4 \times 10^5$ S/m) for the metallic wall boundary conditions. Values in square brackets were observed experimentally using a network analyzer. A few of the resonances (the TE$_{122}$, TE$_{123}$, TE$_{132}$ and the TE$_{134}$) overlap strongly with other modes preventing accurate $Q$ measurements. Those values are given as lower limits.}
\begin{tabular}{|l||l|l|l|}
\hline 
Cavity Mode & $f_o$ (GHz) & $Q$ &$\chi $ ($10^6$) m$^{-3}$ \tabularnewline
\hline
\hline 
TE$_{111}$ &8.17 [8.085] & 190 [300] & 0.20\tabularnewline
TE$_{121}$ &21.8 [21.588] & 1460 [1850] & 0.83\tabularnewline
TE$_{122}$ &23.9 [23.680]& 1280 [$>$430] & 0.61\tabularnewline
TE$_{123}$ & 25.5 [25.191]& 900 [$>$570] & 0.50 \tabularnewline
TE$_{131}$ & 34.2 [33.856]& 2200 [2550] & 1.64 \tabularnewline
TE$_{132}$ & 36.5 [36.605]& 1860 [$>$1220]& 1.22  \tabularnewline
TE$_{133}$ & 38.5 [38.222]& 1500 [1737] & 1.07 \tabularnewline
TE$_{134}$ & 40.3 [40.114]& 1300 [-] & 0.83 \tabularnewline
\hline
\end{tabular}
\end{table}

\subsection{Experimental Characterization}

\begin{figure}
\includegraphics[width=\columnwidth]{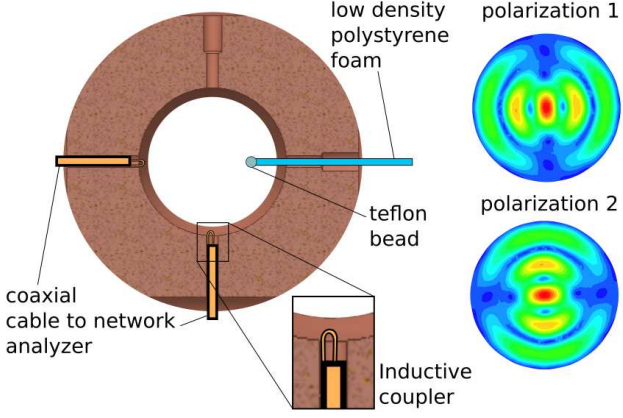}
\caption{\label{fig:VNA} A cavity perturbation technique, together with a network analyzer, allows measurement of resonant frequencies, quality factors, and electric field patterns. Simulated field patterns in the central plane of the cavity are shown for the two polarizations of the TE$_{131}$.}%
\end{figure}

Small holes drilled into the center of the bulge allow inductive loop coupling between the various cavity modes and a network analyzer (see figure \ref{fig:VNA}). The reflection ($S_{11}$) and transmission ($S_{21}$) parameters allow measurement of the cavity resonance frequencies and quality factors. 

\begin{figure}
\includegraphics[width=\columnwidth]{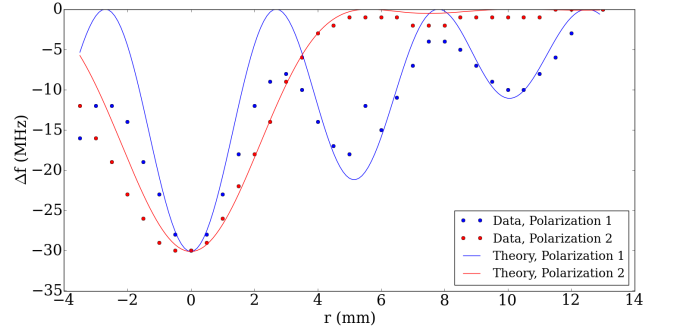}
\caption{\label{fig:FillFactor} The shift in resonant frequency due to a small teflon bead at position $r$ shows the electric field structure of two polarizations of the TE$_{131}$.}%
\end{figure}

By moving a small teflon bead radially throughout the cavity, electric field patterns can be mapped by measurement of the shift in resonant frequency (see figure \ref{fig:FillFactor}). The resonance shift can be used to provide an estimate of the fill factor, $\chi$.  In the limit of a spherical bead, a cavity perturbation analysis \citep{Waldron:2} gives 

\begin{equation}
\frac{ \delta \omega}{\omega_o} = \frac{-3(\epsilon -1)V_1 }{2(\epsilon +2)} \frac{E^2(\vec{x})}{\int_V \! E^2(\vec{x'}) \, \mathrm{d}x'^3 }
\end{equation}

where $\delta \omega$ is the shift in resonance frequency due to the presence of the dielectric, $\omega_o$ is the unperturbed resonance frequency, $V_1$ is the volume and $\epsilon$ is the relative permittivity of the perturbing spherical bead. The fill factor is then

\begin{equation}
\chi = -\frac{ \delta \omega}{\omega_o} \frac{2(\epsilon +2)}{3(\epsilon -1)V_1}
\end{equation}

\subsection{Surface preparations}
\label{sec:surface}

This section describes surface preparations used to lower the cavity $Q$ in an attempt to raise the maximum number of particles our cavity can cool. For an \textit{open} cavity of the type considered here, lowering the $Q$ by means other than changing the surface resistance of the walls will almost certainly \textit{scatter} microwaves out of the resonant mode. For cavity cooling, this is counter-productive.

\begin{figure}
\includegraphics[width=\columnwidth]{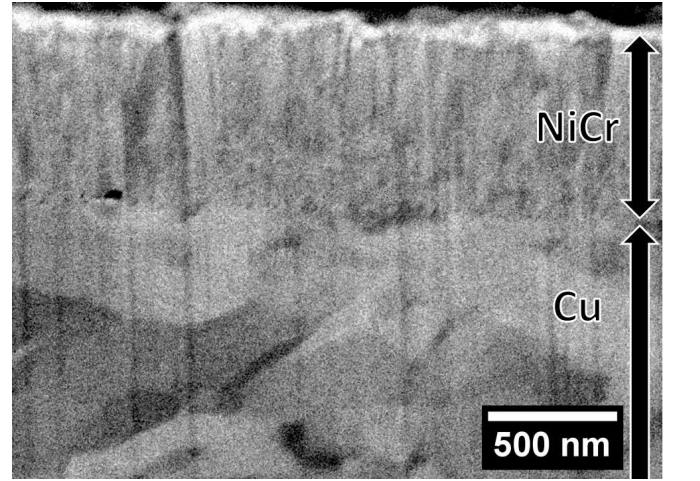}
\caption{\label{fig:SEM} Scanning electron micrograph of the electrodeposited NiCr film cross section after ion milling.}%
\end{figure}

To resistively lower the $Q$, the inner face of the copper resonator was coated with a thin alloy. Nickel-chromium was galvanostatically electroplated onto the inner face of each copper electrode using a previously described process\citep{Tharamani} modified to produce a uniformly thick film with a bright finish. A symmetrical platinum black anode with a surface area approximately twenty times that of the cathode was positioned in the centre of the cavity. Heated plating bath was continuously pumped through the cavity during the deposition on a recirculating loop to minimize porosity induced by adhesion of electrolytically produced hydrogen bubbles on the cathode. Focused ion beam milling and scanning electron microscopy showed the resulting film was 0.7$\mu$m thick, with grains less than 20nm in size (figure \ref{fig:SEM}). The composition of the film was 94$\%$ Ni and 6$\%$ Cr, determined by inductively coupled plasma emission spectroscopy. The high nickel content provides excellent adhesion to the copper substrate and a range of other metals. The current efficiency at 50mA/cm$^2$ was approximately 40$\%$. 

In order to characterize the surface resistance of this alloy on a copper substrate, multiple test cavities were coated using the above process. The cavity geometries tested were a right circular cylindrical and a coaxial cavity, both of which were closed (ie: $Q_{leakage}=\infty$). Measurements of these cavity $Q$'s showed no temperature dependence, as expected for an alloy film. However these $Q$'s indicate that a $\sim1 \mu$m  layer of this alloy on a copper substrate has an effective surface resistance of $R_s \sim$ 0.1 $\Omega/\square$ in the 3 - 18 GHz region. This is higher than can be explained by evaluation of $R_s = \int \! \sigma(x)|E(x)|^2 \, \mathrm{d}x$, where $E(x)$ is the electric field profile inside a semi-infinite metallic slab resulting from unit $H(x=0)$. Continuity of both the $E$ and $H$ fields at the interface between nichrome and copper (where the conductivity, $\sigma(x)$, changes suddenly) is enforced in the functional form of $E(x)$.

Measurements of the bulge TE$_{131}$ $Q$ before and after alloy deposition, together with the simulated $Q_{leakage}$, provide an estimate of the surface resistance of the layered structure as 
\begin{align}
R_{s} =& \frac{Q_{Cu}(Q_{leakage} - Q_{alloy}) }{Q_{alloy}(Q_{leakage} - Q_{Cu})}R_{s,Cu} \\
 \simeq& 2.5 \, R_{s,Cu}.
\end{align}
This is again higher than can be explained using the above model which would suggest that $R_s$ can only be significantly increased if the alloy thickness is of order a skin depth ($\sim 3 \mu$m for nichrome at 30 GHz). In spite of these anomalies, we observe a reduction in $Q$ by a factor of $\sim$2 after applying an alloy layer which is apparently only 0.7$\mu$m thick. Although currently under investigation, the discrepancy between the expected and the apparent properties of the nichrome films is taken to be a material science issue.  

On top of the base nichrome layer, we apply a thin colloidal paste of conductive graphite\citep{EMS}. Static charges on the electrodes have the potential to break the cylindrical symmetry of the Penning trap and induce dichotron instabilities in the plasma \citep{White}. The graphite is applied as an anti-static shield layer to prevent this instability and allow charges which might otherwise accumulate on surface oxide layers to be conducted away. Typically, gold is electroplated onto Penning trap electrodes for this purpose. Here, however, a gold layer would also shield cavity modes from resistive losses in the underlying nichrome. The surface resistance of a graphite film on a glass rod was measured at both 300 K and 4 K and found to be 24 k$\Omega/\square$ and 89 k$\Omega/\square$ $\pm$ 10 $\%$ respectively. This film is effectively in parallel with the surface resistance of the nichrome and therefore produces a negligible effect on the cavity $Q$. This has been verified experimentally.


%
%

%

\begin{acknowledgments}
We gratefully acknowledge Alex Povilus, Eric Hunter and Joel Fajans at UC 
Berkeley for their steadfast interest and assistance in the bulge cavity 
project. WNH also acknowledges Will Bertsche of Manchester University  for 
his early encouragement of the project. Paul Lacarriere and Mohammad Ashkezari are thanked for performing initial microwave simulations on related apparatus. This work was supported by the Natural Sciences and Engineering Research Council of Canada (NSERC).
\end{acknowledgments}

%

\end{document}